\documentclass{article}
  \title{The masses of the mesons and baryons. \\
         Part  I.  The integer multiple rule}
  \author{E.~L. Koschmieder \\
  {\small Center for Statistical Mechanics,  The University of Texas at
Austin,} \\ {\small Austin,  TX 78712,  USA}}
\date{\today}
\begin{document}    
\twocolumn[\maketitle
\begin{center} 
\parbox{4.3in}{\small
From the well-known decays of the particles follows that the
mesons and baryons consist of a $\gamma$-branch and a neutrino branch.
From the well-known masses of the particles follows that the
masses of the $\gamma$-branch particles are integer multiples of the 
mass of the $\pi^0$ meson, within 3\%,  in spite of differences 
in spin,  isospin,  strangeness and charm. The average factor in front
of the integer multiples of $m(\pi^0)$ of the $\gamma$-branch particles 
is $1.0073 \pm 0.0184$. The masses of the $\nu$-branch particles are 
integer multiples of the mass of the $\pi^\pm$ mesons,  times a factor 
$0.86 \pm 0.02$. The existence of the integer multiple rule can be 
verified from the Particle Physics Summary using a calculator.}
\vspace*{3ex} \end{center}]

\section{The  spectrum  of  the  masses  of  the  particles}

The masses of the elementary particles are the best-known and most
characteristic property of the particles.  It seems to be important for
the theoretical explanation of the masses of the particles to find a
simple relationship between the different masses, as the formula for the
Balmer series was important for the explanation of the spectrum of
hydrogen.  We will limit attention here to the mesons and baryons all of
which are unstable, but for the proton.  However, the lifetime of the
mesons and baryons is so long as compared to the period of the basic
frequency $\nu =mc^2 /h$ that the mesons and baryons have often been
categorized as ``stable'' or ``elementary'' particles.  The masses of
the so-called ``stable'' mesons and baryons are given in the ``Particle
Physics Summary'' [1], and are reproduced with other data in Tables
I,II.

It is obvious that any attempt to explain the masses of the elementary
particles should begin with the particles that are affected by the
fewest parameters.  These are certainly the particles without isospin
($I=0$) and without spin ($J=0$), but also with strangeness $S=0$, and
charm $C=0$.  Looking at the particles with $I,J,S,C=0$ it is startling
to find that their masses are quite close to integer multiples of the
mass of the $\pi^0$ meson.  It is $m (\eta ) = (1.0140 \pm 0.0003 )
\cdot 4m (\pi^0 )$, and $m (\eta^\prime ) = (1.0137 \pm 0.00015) \cdot
7m(\pi^0)$.  We also note that the average mass ratios $[m(\eta
)/m(\pi^0) + m(\eta )/m(\pi^+ )]/2 =3.9892 =0.9973 \cdot 4$, and
$[m(\eta^\prime ) /m(\pi^0 ) + m(\eta^\prime )/m(\pi^+ )]/2 =6.9791
=0.9970 \cdot 7$ are good approximations to the integers 4 and 7. Three
particles seem hardly to be sufficient to establish a rule.  However, if
we look a little further we find that $m(\Lambda ) =1.0332 \cdot 8
m(\pi^0 )$ or $m(\Lambda ) =1.0190 \cdot 2 m(\eta )$.  We note that the
$\Lambda$ particle has spin $\frac{1}{2}$, not spin 0 as the $\pi^0$,
$\eta$, $\eta^\prime$, mesons.  Nevertheless, the mass of $\Lambda$ is
close to $8m(\pi^0)$.  Furthermore we have $m(\Sigma^0) =0.9817 \cdot 9
m(\pi^0)$, $m(\Xi^0 ) =0.9742\cdot 10 m(\pi^0)$, $m(\Omega^- ) =
1.0325 \cdot 12 m(\pi^0) =1.0183\cdot 3 m(\eta )$, ($\Omega^-$ is
charged and has spin $\frac{3}{2}$).  Finally the masses of the charmed
baryons are $m(\Lambda_c^+) =0.9958 \cdot 17 m(\pi^0) =1.0232 \cdot 16
m(\pi^+ ) =1.024\cdot 2 m(\Lambda )$, $m(\Sigma_c^0 )=1.0093\cdot 18
m(\pi^0)$, $m(\Xi_c^0) =1.0167\cdot 18 m(\pi^0 )$, and $m(\Omega_c^0
)=1.0017\cdot 20 m(\pi^0)$.  Now we seem to have enough material to
formulate the integer multiple rule, according to which the masses of
the $\eta$, $\eta^\prime$, $\Lambda$, $\Sigma^0$, $\Xi^0$, $\Omega^-$,
$\Lambda_c^+$, $\Sigma_c^0$, $\Xi_c^0$ and $\Omega_c^0$ particles are,
in a first approximation, integer multiples of the mass of the $\pi^0$
meson, although some of the particles have spin, and may also have
charge as well as strangeness and charm.  A consequence of the integer
multiple rule must be that the ratio of any meson or baryon mass divided
by the mass of another meson or baryon is equal to the ratio of two
integer numbers.  And indeed, for example, $m(\eta )/m(\pi^0 )$ is
practically two times (exactly $0.9950\cdot 2$) the ratio $m(\Lambda
)/m(\eta )$, there is also the ratio $m(\Omega^- )/m(\Lambda
)=0.9993\cdot \frac{3}{2} =0.9993\cdot 1.5$.  We have furthermore the
ratios $m(\Lambda )/m(\eta )=1.019\cdot 2$, $m(\Omega^- )/m(\eta )
=1.018\cdot 3$, $m(\Lambda_c^+ ) /m(\Lambda )= 1.02399\cdot 2$,
$m(\Sigma_c^0 ) /m(\Sigma^0 ) =1.0281 \cdot 2$, and $m(\Omega_c^0
)/m(\Xi^0 ) =1.0282 \cdot 2$.

We will call, for reasons to be explained later, the particles discussed
above, which follow in a first approximation the integer multiple rule,
the $\gamma$-branch of the particle spectrum.  The mass ratios of these
particles are listed in Table I. The deviation of the mass ratios from
exact integer multiples of $m(\pi^0)$ is at most 3.3\%, the average of
the factors in front of the integer multiples of $m(\pi^0)$ of the ten
$\gamma$-branch particles in Table I is $1.0073 \pm 0.0184$.  From a
least square analysis follows that the masses of the eleven particles on
Table I obey the formula $m= 1.0059 N +0.0074$ with a correlation
coefficient $r=0.999$.  The consequences of the combination of spin,
isospin, strangeness and charm are difficult to assess.  Nevertheless,
even the combination of four different parameters does not change the
integer multiple rule by more than 3.3\%. To put this into perspective 
we note that the masses of the $\pi^\pm$ mesons and the $\pi^0$ meson
differ already by 3.4\%.

Spin seems to have a profound effect on the mass of a particle.
Changing the spin from zero for the $\pi^0$, $\eta$, $\eta^\prime$,
mesons to spin $\frac{1}{2}$ for the $\Lambda$ baryon is accompanied by
a mass twice the mass of the $\eta$ meson, it is $m(\Lambda )
=1.0190\cdot 2 m(\eta )$.  The isospins of $\eta$ and $\Lambda$ are both
zero.  The change of $S$ and the baryon number $B$ which accompany the
formation of $\Lambda$ seems to have little effect on the mass of the
$\Lambda$ particle and the other baryons, as follows from the masses of
the baryons whose $S$ changes from $-1$ to $-3$.  We find it most
interesting that spin $\frac{3}{2}$ is accompanied with a mass three
times $m(\eta )$, the $\Omega^-$ particle; whereas spin $\frac{1}{2}$ is
accompanied by a mass two times $m(\eta )$, the $\Lambda$ particle.

Searching for what the $\pi^0$, $\eta$, $\eta^\prime$, $\Lambda$,
$\Sigma^0$, $\Xi^0$, $\Omega^-$ particles have else in common, we find
that the principal decays (decays with a fraction $> 1$\%) of these
particles, as listed in Table I, involve primarily $\gamma$'s, the
characteristic case is $\pi^0 \rightarrow \gamma \gamma $ (98.8\%). The
next most frequent decay product of the heavier particles of the
$\gamma$-branch are $\pi^0$ mesons which again decay into $\gamma
\gamma$.  To describe the decays in another way, the principal decays of
the particles listed above take place \textit{always without the
emission of neutrinos}; see Table I. There the decays and the fractions
of the principal decay modes are listed, as they are given in the
Particle Physics Summary.  We cannot consider decays with fractions
$<1$\%. We will refer to the particles whose masses are approximately
integer multiples of the mass of the $\pi^0$ meson, and which decay
without the emission of neutrinos, as the $\gamma$-branch of the
particle spectrum.

To summarize the facts concerning the $\gamma$-branch.  Within about
3\%\ the masses of the particles of the $\gamma$-branch are integer
multiples (namely 4, 7, 8, 9, 10, 12, and even 17, 18, 20) of the mass
of the $\pi^0$ meson.  It is improbable that nine particles have masses
so close to integer multiples of $m(\pi^0)$ if there is no correlation
between them and the $\pi^0$ meson.  It has, on the other hand, been
argued that the integer multiple rule is a numerical coincidence.  But
the probability that the mass ratios of the $\gamma$-branch fall by
coincidence on integer numbers between 1 and 20 instead on all possible
percentage values between 1 and 20 is smaller than $10^{-20}$, i.e.,
nonexistent.  The integer multiple rule is not affected by more than
3\%\ by the spin, the isospin, the strangeness, and by charm.  The
integer multiple rule seems even to apply to the $\Omega^-$ and
$\Lambda_c^+$ particles, although they are charged.  In order for the
integer multiple rule to be valid the deviation of the ratio $m/m(\pi^0
)$ from an integer number must be smaller than $1/2N$, where $N$
is the integer number closest to the actual ratio $m/m(\pi^0 )$.  That
means that the permissible deviation decreases rapidly with $N$.  All
particles of the $\gamma$-branch have deviations smaller than
$1/2N$.

The remainder of the stable mesons and baryons are the $\pi^\pm$,
$K^{\pm ,0}$, $p$, $n$, $D^{\pm ,0}$ and $D_S^\pm$ particles.  These are
in general charged, exempting the $K^0$ and $D^0$ mesons and the neutron
$n$, in contrast to the particles of the $\gamma$-branch, which are in
general neutral.  It does not make a significant difference whether one
considers the mass of a particular charged or neutral particle.  After
the $\pi$ mesons, the largest mass difference between charged and
neutral particles is that of the $K$ mesons (0.81\%), and thereafter
all mass differences between charged and neutral particles are $< 
0.5$\%.  The integer multiple rule does not immediately apply to the
masses of the charged particles if $m(\pi^\pm )$ (or $m(\pi^0)$) is used
as reference, because $m(K^\pm ) =0.8843 \cdot 4 m(\pi^\pm )$.  $0.8843
\cdot 4$ is far from integer.  Since the masses of the $\pi^0$ meson and
the $\pi^\pm$ meson differ by only 3.4\%\ it has been argued that the
$\pi^\pm$ mesons are, but for the isospin, the same particles as the
$\pi^0$ mesons, and that therefore the $\pi^\pm$ cannot start another
particle branch.  However, this argument is not supported by the
completely different decays of the $\pi^0$ mesons and the $\pi^\pm$
mesons.  The $\pi^0$ meson decays almost exclusively into $\gamma
\gamma$ (98.8\%), whereas the $\pi^\pm$ mesons decay practically
exclusively into $\mu$-mesons and neutrinos, e.g.\ $\pi^+ \rightarrow
\mu^+ + \nu_\mu$ (99.987\%). Furthermore, the lifetimes of the $\pi^0$
and the $\pi^\pm$ mesons differ by nine orders of magnitude, being $\tau
(\pi^0 ) =8.4\cdot 10^{-17}$ sec \textit{versus} $\tau (\pi^\pm )
=2.6\cdot 10^{-8}$ sec.

If we make the $\pi^\pm$ mesons the reference particles of the 
$\nu$-branch, then we must multiply the mass ratios $m/m(\pi^\pm )$ of
the above listed particles with a factor $0.861 \pm 0.022$, as follows
from the mass ratios listed on Table II.  The integer multiple rule may,
however, apply directly if one makes $m(K^\pm )$ the reference for
masses larger than $m(K^\pm )$.  The mass of the proton is $0.9503\cdot
2 m(K^\pm )$, which is only a fair approximation to an integer multiple.
There are, on the other hand, outright integer multiples in $m(D^\pm )
=0.9961\cdot 2 m(p)$, and in $m(D_S^\pm ) =0.9968\cdot 4 m (K^\pm )$.
We note that the spin $\frac{1}{2}$ of the proton is associated with a
mass twice the mass of the spinless $K$ meson, just as it was with the
spin of the $\Lambda$ particle, which is associated with a mass twice
the mass of the spinless $\eta$ meson.  We note further that the spin of
the $D^\pm$ mesons, whose mass is nearly $2m(p)$, is zero, whereas the
spin of the proton is $\frac{1}{2}$.  It appears that the superposition
of two particles of the same mass and with spin $\frac{1}{2}$ can cancel
the spin.  On the other hand, it appears that the superposition of two
particles of equal mass without spin can create a particle with spin
$\frac{1}{2}$.

Contrary to the particles of the $\gamma$-branch, the charged particles
decay preferentially with the emission of neutrinos, the foremost
example is $\pi^+ \rightarrow \mu^+ + \nu_\mu$ with a fraction of
99.987\%. Neutrinos characterize the weak interaction.  We will refer
to the charged particles listed in Table II as the \textit{neutrino
branch} ($\nu$-branch) of the particle spectrum.  We emphasize that a
weak decay of the particles of the $\nu$-branch is by no means
guaranteed, the proton is stable and there are decays as, e.g., $K^+
\rightarrow \pi^+ \pi^- \pi^+$ (5.59\%), but the subsequent decays of
the $\pi^\pm$ mesons lead, on the other hand, to neutrinos and $e^\pm$.
There is also the $K^0$ particle, which poses a problem because the
principal primary decays of $K^0_S$ take place without the emission of
neutrinos, but many of the secondary decays emit neutrinos.  On the
other hand, 2/3 of the primary decays of $K^0_L$ emit neutrinos.  For
comparison 63.6\%\ of $K^+$ decay into $\mu^+ + \nu_\mu$.  The decays of
the particles in the $\nu$-branch follow a mixed rule, either weak or
electromagnetic.

To summarize the facts concerning the $\nu$-branch of the mesons and
baryons.  The masses of these particles seem to follow an integer
multiple rule if one uses the $K^\pm$ meson as reference, however the
masses are not integer multiples of the $\pi^\pm$ mesons but share a
common factor $0.86 \pm 0.02$.

We do not want to discuss here the bottom mesons, and the bottom baryon.
It is easy to associate the masses of the bottom mesons with integer
multiples of the $\pi^0$ meson.  For example, $m(B^0 )=1.0029\cdot 39
m(\pi^0 )$ and $m(B_S^0 )=0.9945\cdot 40 m(\pi^0 )$ or $1.0199\cdot 39
m(\pi^0 )$.  The latter numbers show the difficulty in identifying the
proper multiple.  In this case the difference of the integer multiples
is one out of forty or 2.5\%, which is too small in order to make a
proper identification of the multiple.  There are, however, the
$c\overline{c}$ mesons $\eta_c$ and $\chi_{c0}$, with $I,J=0$, whose
masses are $1.0035\cdot 22 m(\pi^0 )$ and $1.0120\cdot 25 m(\pi^0 )$,
which therefore fall into the range of the masses of the $\gamma$-branch
considered here, and which are good approximations to integer multiples
of $m(\pi^0 )$.

{\small
\begin{table}
\caption{The $\gamma$-branch of the particle spectrum.}
\begin{center} \begin{tabular}{@{}lccccc@{}} \hline \hline
& $m/m(\pi^0 )$ & multiples & decay & fraction & spin \\
&               &           &       & (\%)    & \\ \hline
$\pi^0$ & 1 & $1\cdot \pi^0$ & $\gamma \gamma$ & 98.798 & 0 \\
        &   &                & $e^+ e^- \gamma$ & 1.198 & \\[1ex]
$\eta$ & 4.0559 & $1.0140\cdot 4\pi^0$ & $\gamma \gamma$ & 39.25 & 0 \\
       &        &                      & $3\pi^0$        & 32.1 & \\
       &        &                      & $\pi^+ \pi^- \pi^0$& 23.1 & \\
       &        &                      & $\pi^+ \pi^- \gamma$ & 4.78 & 
\\[1ex]
$\eta^\prime$& 7.096 & $1.0137\cdot 7\pi^0$& $\pi^+ \pi^- \eta$ &
             43.7 & 0 \\
       &        &                      & $\rho^0 \gamma$ & 30.2 & \\
       &        &                    & $\pi^0 \pi^0 \eta$ & 20.8 & \\
       &        &                    & $\omega \gamma$ & 3.02 & \\
       &        &                    & $\gamma \gamma$ & 2.12 & \\[1ex]
$\Lambda$& 8.26577 & $1.0332\cdot 8\pi^0$ & $p \pi^-$& 63.9 &     
      $\frac{1}{2}$ \\ 
       &        & $1.0190\cdot 2\eta$& $n\pi^0$&  35.8 & \\[1ex]
$\Sigma^0$& 8.8352 & $0.9817\cdot 9\pi^0$& $\Lambda \gamma$ & 100 &
      $\frac{1}{2}$ \\[1ex]
$\Xi^0$& 9.7417 & $0.9742\cdot 10\pi^0$ & $\Lambda \pi^0$ & 99.54 &
      $\frac{1}{2}$ \\[1ex]
$\Omega^-$ & 12.390 & $1.0326\cdot 12\pi^0$ & $\Lambda K^-$ & 67.8 &
      $\frac{3}{2}$ \\
       &        & $0.9986\cdot 12\pi^-$ & $\Xi^0 \pi^-$ & 23.6 & \\
       &        & $1.0183\cdot 3\eta$ & $\Xi^- \pi^0$ & 8.6 & \\[1ex]
$\Lambda_c^+$& 16.928 & $0.9958\cdot 17\pi^0$ & many & & $\frac{1}{2}$\\
       &        & $0.9630\cdot 17\pi^+$ & & & \\[1ex]
$\Sigma_c^0$ & 18.167 & $1.0093\cdot 18\pi^0$ & $\Lambda_c^+ \pi^- $ &
       $\approx 100$ & $\frac{1}{2}$ \\[1ex]
$\Xi_c^0$& 18.302& $1.0167\cdot 18\pi^0$ & & & $\frac{1}{2}$ \\[1ex]
$\Omega_c^0$& 20.033 & $1.0017\cdot 20\pi^0$ & & & $\frac{1}{2}$ 
\\ \hline \hline
\end{tabular} \end{center} \end{table}}

\section{Summary}

  In spite of differences in charge,  spin, strangeness, and charm
the masses of the stable mesons and baryons of the $\gamma$-branch are, 
within at most 3.3\%, integer multiples of the mass of the $\pi^0$
meson.  Correspondingly, the masses of the particles of the $\nu$-branch
are, after multiplication with a factor $0.86\pm 0.02$, integer
multiples of the mass of the $\pi^\pm$ mesons.  The validity of the
integer multiple rule can easily be verified from the Particle Physics
Summary using a calculator.  The integer multiple rule suggests that the
particles are the result of superpositions of modes and higher modes of
a wave equation.  Such a theory will be presented in a following paper.
\vspace{1ex}

The author is grateful to Professor I.\ Prigogine for his support.
\\[2ex]
REFERENCE \\[1ex]
[1] R. Barnett et al., Rev.\ Mod.\ Phys.\ \textbf{68}, 611 (1996).

\begin{table} 
\caption{The $\nu$-branch of the particle spectrum}
\begin{center}
\begin{tabular*}{3.7in}{@{}l@{\extracolsep{\fill}}cclc@{}}
 \hline \hline
     & $m/m(\pi^+)$ & multiples & decay\footnotemark
     \hfill fraction & spin \\
&              &           &       \hfill (\%) & \\ \hline 
$\pi^\pm$ & 1 & $1\cdot \pi^+$ & $\mu^+ \nu_\mu$ 
      \hfill 99.987 & 0 \\[1ex]
$K^\pm $ & 3.53713 & $0.8843\cdot 4\pi^+$ & $\mu^+ \nu_\mu$ 
      \hfill 63.57 & 0 \\
      &         &                      &$\pi^+ \pi^0$ 
      \hfill 21.16 & \\
      &         &           &$\pi^+ \pi^- \pi^+$ 
      \hfill 5.59 & \\
      &         &           &$\pi^0 \mu^+ \nu_\mu$ 
      \hfill 3.18 & \\
      &         &           &$\pi^0 e^+ \nu_e$ 
      \hfill 4.82 & \\[1ex]
$p^+$ & 6.722595& $0.8403\cdot 8\pi^+$ & stable  & $\frac{1}{2}$ \\ 
      &         & $0.9503\cdot 2K^+$ &  & \\
      &         & $0.9604\cdot 7\pi^+$  & & \\[1ex]
$D^\pm$ & 13.393& $0.8370\cdot 16\pi^+$ & $e^+${\small \it anything} 
      \hfill 17.2 & 0 \\
      &         & $0.9466\cdot 4 K^+$ & $K^-${\small \it anything} 
      \hfill 24.2 & \\
      &         & $0.9961\cdot 2p$ & 
                         $\overline{K}^0${\small \it anything}  & \\
      &         &                  &
                         +$K^0${\small \it anything} 
      \hfill 59 & \\
      &         &        & $\eta$ {\small \it anything} 
      \hfill $< \! 13$ & \\[1ex]
$D_S^\pm$& 14.104& $0.8815\cdot 16\pi^+$ & $K^-${\small \it anything} 
      \hfill 13 & 0 \\
      & & $0.9968\cdot 4 K^+$& $\overline{K}^0${\small \it anything} 
      & \\
      &         &          & + $K^0${\small \it anything} 
      \hfill 39 & \\
      &         &          &  $K^+${\small \it anything} 
      \hfill 20 & \\
      &         &          & $e^+${\small \it anything} 
      \hfill $< \! 20$ & \\
 \hline \hline
\end{tabular*} \end{center}
\mbox{}$^1$ The particles with negative charge have conjugate
charges of the decays listed.
\end{table}
\end{document}